\documentstyle[multicol,epsf,aps]{revtex}
\begin{document}
\draft
\title{Shell model calculations of stellar weak interaction rates: \\
I. Gamow-Teller distributions and spectra of nuclei in the mass range
$A=45-65$} 

\author{E. Caurier$^1$, K. Langanke$^2$, G. Mart\'{\i}nez-Pinedo$^2$,
  and F. Nowacki$^3$ }

\address{$^1$Institut de Recherches Subatomiques
  (IN2P3-CNRS-Universit\'e Louis Pasteur) B\^at.27/1, F-67037
  Strasbourg Cedex~2, France}

\address{$^2$Institute of Physics and Astronomy, University of
  {\AA}rhus, DK-8000 {\AA}rhus C, Denmark}

\address{$^3$Laboratoire de Physique Th\'eorique de Strasbourg, 3-5
  rue de L'Universit\'e, F-67084 Strasbourg Cedex, France}

\date{\today} 

\maketitle

\begin{abstract}
  Electron capture and beta-decay rates on nuclei in the mass range
  $A=45-65$ play an important role in many astrophysical environments.
  The determination of these rates by large-scale shell model
  calculations is desirable, but it requires to reproduce the
  Gamow-Teller strength distributions and spectra of the $pf$ shell
  nuclei.  We show in this paper that large-scale shell model
  calculations, employing a slightly monopole-corrected version of the
  wellknown KB3 interaction, fulfill these necessary requirements.  In
  particular, our calculations reproduce the experimentally available
  GT$_+$ and GT$_-$ strength distributions and the nuclear halflives,
  and describe the nuclear spectra appropriately.
\end{abstract}


\begin{multicols}{2}
 
For many years it has been recognized that nuclear beta-decay and
electron capture are important during the late stages of stellar
evolution. Shortly after the discovery of the Gamow-Teller (GT)
resonance, Bethe {\it et al.} \cite{BBAL} recognized the importance of
this collective nuclear excitation for stellar electron capture.  In
the seminal work by Fuller, Fowler and Newman (usually abbreviated as
FFN, \cite{FFN}) stellar electron capture and beta-decay rates have
been systematically estimated for nuclei in the mass range $A=45-60$
considering two distinct contributions. At first, these authors
estimated the GT contributions to the rates by a parametrization based
on the independent particle model. The rate estimate has then been
completed by experimental data for discrete transitions, whenever
available. Unmeasured allowed GT transitions have been assigned an
empirical value (log ft=5).  One of the important ideas in FFN was to
recognize the role played by the Gamow-Teller resonance in $\beta$
decay. Other than in the laboratory, $\beta$ decay rates under stellar
conditions are significantly increased due to thermal population of
the Gamow-Teller back resonance in the parent nucleus (the GT back
resonance are the states reached by the strong GT transitions in the
inverse process (electron capture) built on the ground and excited
states, see \cite{FFN}) allowing for a transition with a large nuclear
matrix element and increased phase space.  Thus it follows that both,
stellar electron capture and beta decay, are very sensitive to the
distribution of the GT$_+$ strength (in this direction a proton is
changed into a neutron). In the last decade, GT$_+$ strength
distributions on nuclei in the mass range $A=50-65$ have been studied
experimentally via (n,p) charge-exchange reactions at forward angles
and the experimental data
\cite{gtdata1,gtdata2,gtdata3,gtdata4,gtdata5} show that, in contrast
to the independent particle model, the total GT$_+$ strength is
quenched and fragmented over many final states in the daughter nucleus
caused by residual nucleon-nucleon correlations.  Importantly for the
stellar weak reaction rates, the data indicate a systematic
misplacement of the GT centroid adopted in the FFN parametrization.
The need for an improved theoretical description of stellar weak
interaction rates has consequently been realized.

The method of choice for these calculations is the interacting shell
model. Early attempts to improve selected FFN rates
\cite{Aufderheide91,Aufderheide93a,Aufderheide93b,Aufderheide96} have
been performed within the conventional shell model diagonalization
approach, however, in strongly restricted model spaces and with
residual interactions, which did neither reproduce the quenching nor
the position of the GT strength distributions well enough to give
these approaches the necessary predictive power.  One of the
limitations of these shell model calculations (the strongly
restrictive model spaces) has been overcome recently by decisive
progress achieved in shell model technologies. At first, the
development of the shell model Monte Carlo technique (SMMC)
\cite{Johnson,Koonin97} allows now calculations of the GT strength
distributions in the complete $pf$ shell. But also with
state-of-the-art diagonalization approaches, combined with modern
computer technologies, shell model calculations can be performed on a
truncation level which guarantees basically complete and converged
calculations of the GT strength distributions for $pf$ shell nuclei.
Although the first shell model estimates of stellar electron capture
rates has been presented on the basis of the SMMC approach
\cite{Dean98}, we note that the diagonalization method is likely
better suited for the calculation of stellar weak interaction rates.
The reason is that the diagonalization approach allows for detailed
spectroscopy, while the SMMC model yields only an ``averaged'' GT
strength distribution which introduces some inaccuracies into the rate
calculations. Furthermore, SMMC calculations for odd-$A$ and odd-odd
nuclei can only be performed at temperatures $T> 0.8$ MeV, which is in
access of the temperatures encountered at the onset of the core
collapse of a massive star.

This paper is concerned with the solution of the second problem,
encountered in the early studies: the lack of an appropriate residual
interaction which can be used to calculate the stellar weak
interaction rates within the entire mass range $A=45-65$.  To do so,
we have to recall the important ingredients necessary to estimate
stellar weak interaction rates; from these ingredients we will derive
requirements to be fulfilled by shell model calculations to have
predictive power in estimating these rates.  At first, we stress that
the GT$_+$ strength distribution directly determines the electron
capture rates and, via the back-resonance, also indirectly contributes
significantly to or even dominates the beta-decay rates.  However, the
beta-decay rates are not entirely given by the back-resonance
contribution, as the ground state and other low-lying states can also
contribute. These contributions are determined mainly by the low-lying
part of the respective GT$_-$ strength distribution (in which a
neutron is changed into a proton and which can be experimentally
determined via (p,n) charge-exchange reactions).  Finally, under
stellar conditions the weak processes occur at finite temperature and
hence several thermally populated states at modest temperatures (in
beta decay supplemented by the back-resonances) will potentially
contribute to the stellar rates.  To be applicable to calculating
stellar weak interaction rates the shell model calculations (model
space plus residual interaction) should reproduce most notably the
available GT$_+$ strength distributions, but they should also give a
fair account of the GT$_-$ strength distributions and the nuclear
spectra.

Why is the original KB3 interaction not suited to calculate the weak
interaction rates in the entire mass range $A=45-65$?  To understand
the problem and to find a possible cure we recall the following
observations.  Shell model calculations, performed in the complete
$pf$ shell and employing the KB3 interaction \cite{Zuker81}, have been
demonstrated to give an excellent description of nuclei at the
beginning of the shell ($A<50$) \cite{Caurier94,Martinez96}. However,
the extension of these complete $0 \hbar\omega$ studies towards
heavier nuclei, performed with the SMMC approach, clearly demonstrated
the decreasing level of accuracy of the KB3 interaction which, for
example, results in an overenhancement of the $N=28$ shell closure,
most notably in $^{56}$Ni \cite{Langanke95}. This finding has recently
been confirmed and explored in more details on the basis of shell
model diagonalization calculations \cite{Nowacki98}. Furthermore it
turns out that the KB3 interaction apparently does not give a
meaningful ordering of the single particle energies in the upper $pf$
shell, where the $f_{5/2}$ orbit is pushed up relative to the $p$
orbits \cite{Ormand}.

To understand the cure of the problem, let us remember that the KB3
interaction is nothing else but the original Kuo-Brown interaction,
however, with a slight, but important correction of the monopole part
\cite{Zuker81}.  Due to numerical restrictions at the time, the fixing
of the monopole part had to be limited to pure $f_{7/2}$ nuclei,
leaving the non-$f_{7/2}$ monopoles essentially undetermined. In fact,
it turns out that the quasiparticle gap in $^{56}$Ni is too strong
resulting in a relative underbinding of nuclei with neutron or proton
numbers larger than 28.

The two failures of the KB3 interaction mentioned above suggest that
i) the $f_{7/2}$ monopole correction introduced in \cite{Zuker81} has
been slightly too large and ii) the $f_{5/2}$ monopole term should be
lowered with respect to the $p$ orbits. This leads us to the following
modifications of the centroids $V^T_{ij}$ between any two shells
($ij)$ in the KB3 interaction (for a definition of the centroids the
reader is refered to \cite{Caurier94}): $V^T_{7/2,r}$ with
(r=3/2,5/2,1/2) is lowered by 0.1~MeV; $V^T_{3/2,3/2}$,
$V^T_{3/2,1/2}$, and $V^T_{1/2,1/2}$ are all raised by 0.1~MeV; and
$V^T_{5/2,5/2}$, $V^T_{5/2,3/2}$, and $V^T_{5/2,1/2}$ are all lowered
by 0.2~MeV.  These corrections apply to both isospins $T=0,1$.

Let us stress here that we have not aimed to optimize the monopole
corrections. They are rather the minimal ones which allow a fair
reproduction of the GT strength distributions and halflives, but for
more detailed spectroscopic studies it is probably required to correct
some minor deficiencies seen in occasional inversions of levels at the
higher mass numbers (see below).  There are ongoing efforts to
overcome these minor inefficiencies and to derive a `final' residual
interaction for the $pf$ shell.  However, as we will demonstrate now,
our interaction is already a well suited tool to calculate the stellar
weak interaction rates on the basis of large-scale shell model
calculations. We like to stress that slight energy misplacements in
the level scheme do not matter, as in the calculation of the rates
\cite{Langanke99} the shell model energies will be replaced by
experimental data.

We will now present the results of large-scale shell model
calculations performed with our interaction defined above. Having our
ultimate goal in mind - a compilation of stellar electron capture and
beta-decay rates for more than 100 nuclei in the mass range $A=45-65$
\cite{Langanke99} - we had to find a balance between numerical
accuracy and computational feasibility. Thus we have not always
performed the calculations at the highest truncation level achievable
on modern computers (note that the rate calculations require the
evaluation of the GT strengths for many levels per nucleus).  However,
at the chosen level of truncation involving typically 10 million
configurations or more, the GT strength distribution is virtually
converged for the nuclei studied.  We also mention that the GT
strengths distributions have always been calculated in truncated model
spaces which fulfill the Ikeda sum rule.  Finally, as $0\hbar\omega$
shell model calculations, i.e.  calculations performed in one major
shell, overestimate the experimental GT strength by a universal factor
\cite{Wildenthal,Langanke95,Martinez96}, we have scaled our GT
strength distribution by this factor, $(0.74)^2$.  To allow the reader
the judgment of quality achieved in the shell model calculations
underlying the stellar weak interaction rate compilation the results
presented below have been calculated within the same truncation scheme
as the compilation.  Some electron capture and beta decay rates, based
on our shell model studies, have already been reported in
\cite{Langanke98a,Langanke98b,Martinez99}.

For the mass range $A=45-65$ of interest here, GT$_+$ strength
distributions have been determined from (n,p) charge-exchange
experiments for the even-even nuclei $^{54,56}$Fe, $^{58,60,62,64}$Ni
and the odd-$A$ nuclei $^{51}$V, $^{55}$Mn, $^{59}$Co
\cite{gtdata1,gtdata2,gtdata3,gtdata4,gtdata5}.  In Table I we have
listed the experimental and calculated total GT$_+$ strengths.  The
overall agreement is very satisfying.  The level of convergence
achieved with the chosen truncation level can be estimated to be of
the order $10\%$ or better. To check this we have performed a study of
$^{58}$Ni in a model space in which a maximum of 6 particles were
allowed to be excited from the $f_{7/2}$ shell to the rest of the $pf$
shell in the final nucleus; the value quoted in Table I has been
calculated for a maximum of 5 particles. With the increased truncation
level the total GT$_+$ strength reduces to 4.1 units, a decrease of
about $7\%$.  We also note that the data for the odd-A nuclei
($^{51}$V, $^{55}$Mn, $^{59}$Co) show additional GT$_+$ strength at
higher excitation energies between $E_x=8-12$ MeV, which has not been
considered in the values quoted in Table I. Aufderheide {\it al.} have
analyzed the data for $^{51}$V and $^{59}$Co upto $E_x=12$ MeV and
quote GT$_+$ values of 1.5$\pm$0.2 and 2.4$\pm$0.3 for these two
nuclei \cite{Aufderheide93a}, in very close agreement with the shell
model results.  We also note that, compared to experiment, the
calculated GT$_+$ strength for $^{64}$Ni is somewhat too small; a
tendency which might already been indicated in $^{62}$Ni. It is
conceivable that a better description of these nuclei require the
inclusion of the $g_{9/2}$ orbital in the model space, a venture which
is beyond the scope of the present study.

\begin{figure}
  \begin{center}
    \leavevmode
    \epsfxsize=0.8\columnwidth
    \epsffile{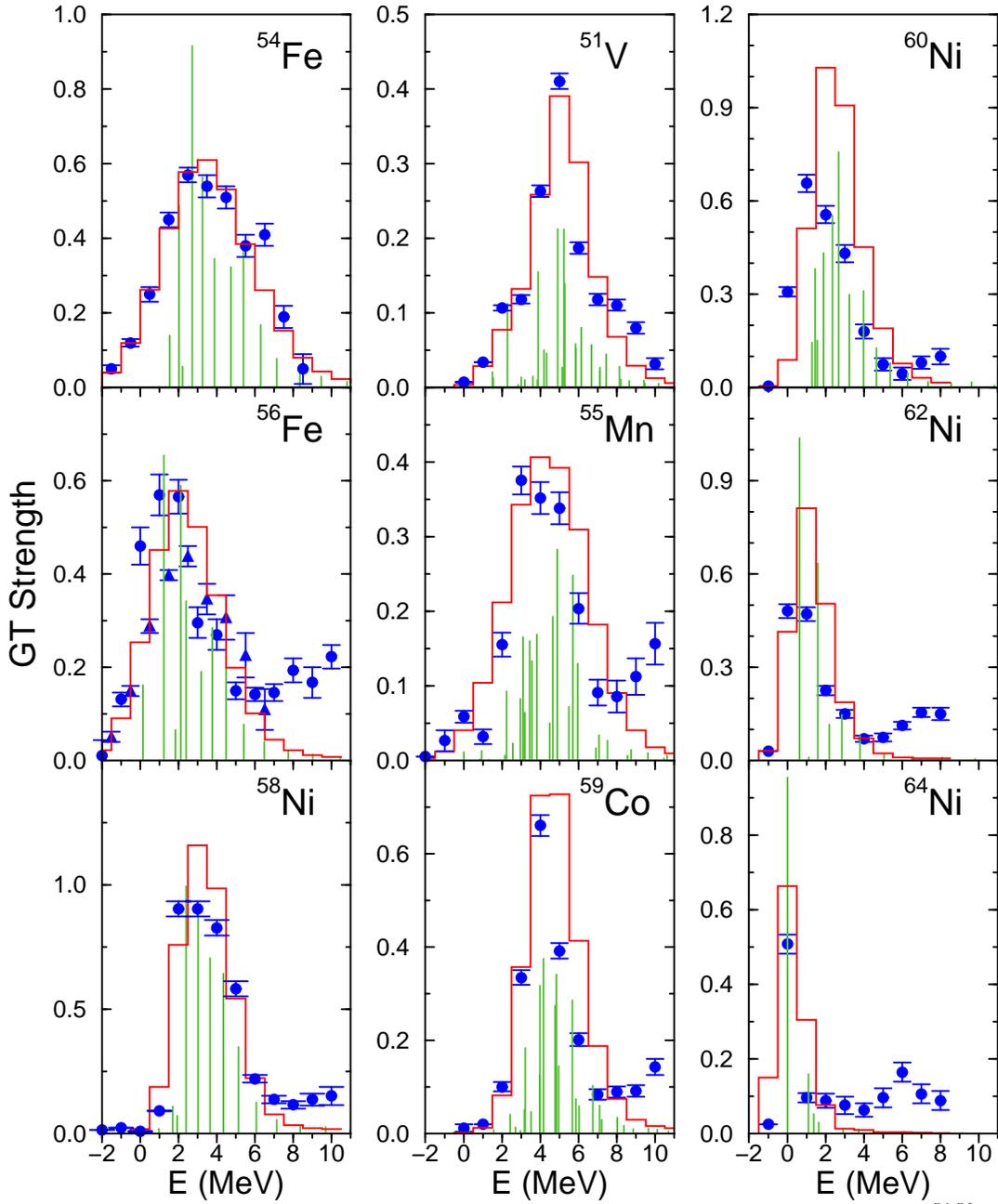}
    \caption{Comparison of GT$_+$ data with our calculated strength
      distributions for the even-even nuclei $^{54,56}$Fe and
      $^{58-64}$Ni and the odd-$A$ nuclei $^{51}$V, $^{55}$Mn and
      $^{59}$Co.  The shell model results are shown by discrete spikes
      which have been folded with the experimental energy resolution
      to obtain the plotted histograms.  The data are from
      \protect\cite{gtdata1,gtdata2,gtdata3,gtdata4,gtdata5}.  Note
      that the experimental GT strength distribution for $^{60-64}$Ni
      does not add up to the total GT strength quoted in the same
      reference \protect\cite{gtdata5} (see text).  }
    \label{fig1}
  \end{center}
\end{figure}

For the calculation of the stellar weak interaction rates, a proper
reproduction of the GT strength distribution is more important than of
the total strength, due to a very strong energy dependence of the
phase space factors.  In Fig.~\ref{fig1} we compare our calculated
GT$_+$ distributions with the data. To allow for a meaningful
comparison we have folded the shell model results with Gaussians to
account for the experimental resolution.  The agreement is again
satisfying in all cases.  It is interesting to compare the present
distributions to those obtained for the KB3 interaction within the
SMMC approach, e.g.  Figs. 3 and 4 of Ref. \cite{Radha97}. One clearly
observes that the slight monopole corrections, defined above, result
in a much better reproduction of the data. This is most noticeably
visible for the nickel isotopes. However, we also note that the
calculated strength function for $^{60}$Ni is slightly shifted to
higher energies compared to the data (by about 500 keV).

For the nickel isotopes $^{60,62,64}$Ni we have taken the GT$_+$ data
from Fig.~12 in \cite{gtdata5}. However, summing these data upto 8.5
MeV yields a total GT strength, which is about $20\%$ smaller than the
values cited in \cite{gtdata5} and Table I, and plotted in Fig.~10 of that
reference.

In Table I we also compare the centroids of the calculated GT$_+$
distributions with the GT resonance energy assumed in the FFN
parametrization. We find that for even-even nuclei the GT centroid is
usually at lower excitation energy in the daughter nucleus than
assumed in \cite{FFN}. For odd-A nuclei, Fuller {\it et al.} placed
the GT resonance at lower excitation energies than calculated in our
shell model approaches or observed in the data. For odd-odd nuclei the
shell model centroids \cite{Langanke98b} are also at higher excitation
energies than adopted in \cite{FFN}.  Possible consequences of these
misplacements for the stellar electron capture and beta decay rates
are discussed
in~\cite{Aufderheide96,Langanke98a,Langanke98b,Martinez99}.

\begin{figure}
  \begin{center}
    \leavevmode
    \epsfxsize=0.8\columnwidth
    \epsffile{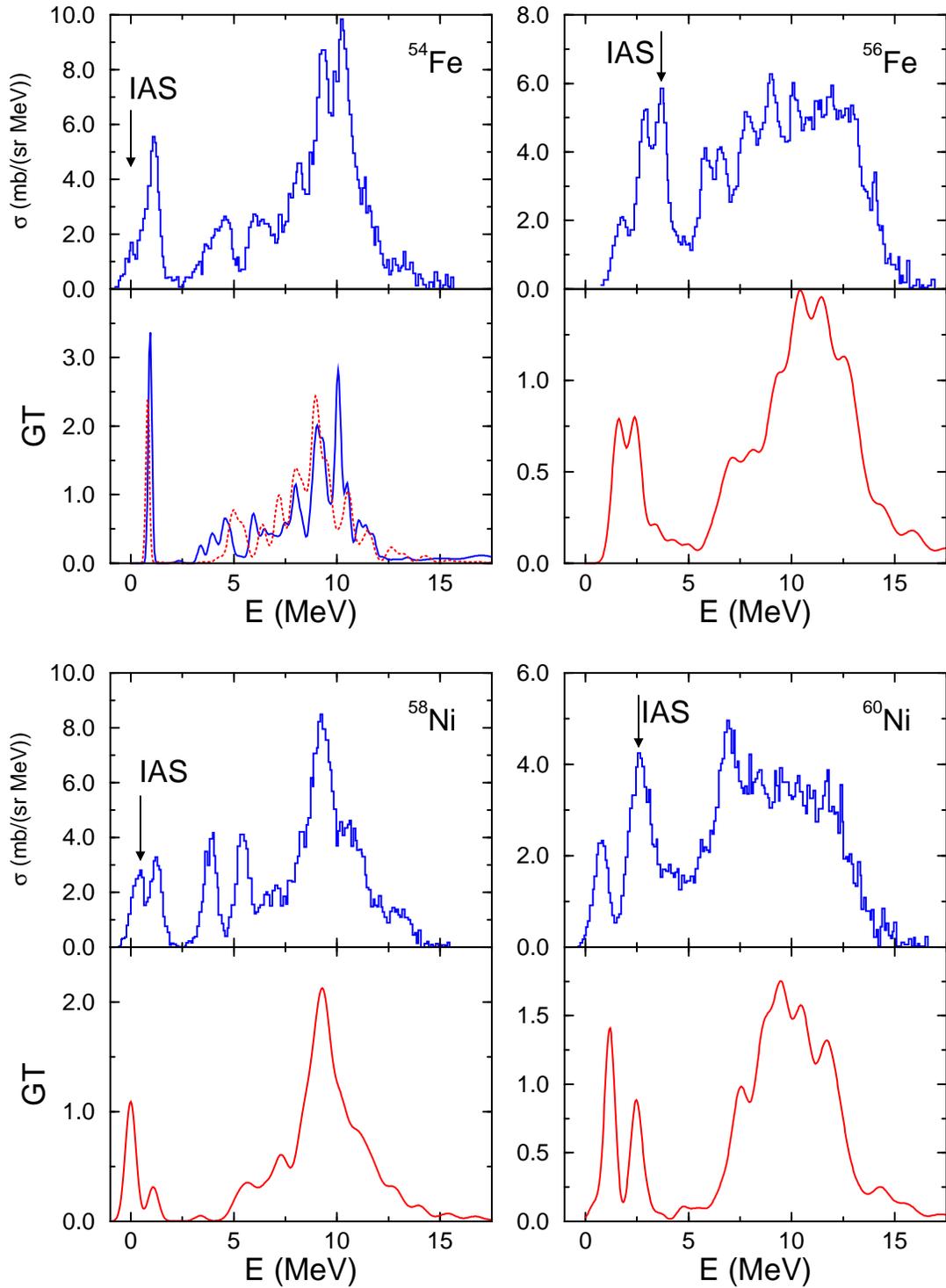}
    \caption{Comparison of (p,n) $L=0$ forward-angle cross section
      data~\protect\cite{Rapaport} (upper panels) with the calculated
      GT$_-$ strength distributions (lower panels) for the nuclei
      $^{54}$Fe, $^{56}$Fe, $^{58}$Ni, and $^{60}$Ni. For $^{54}$Fe
      the solid curve in the lower panel shows the experimental GT$_-$
      data from~\protect\cite{Anderson}, while the present shell model
      results are given by the dotted curve. The Fermi transition to
      the isobaric analog state (IAS) is included in the data (upper
      panels) but not in the calculations.}
    \label{fig2}
  \end{center}    
\end{figure}

Informations about the GT$_-$ strength distribution has been
determined experimentally from (p,n) charge-exchange reaction
measurements for the nuclei $^{54,56}$Fe and $^{58,60}$Ni
\cite{Anderson,Rapaport}. As these Gamow-Teller transitions change a
neutron into a proton, they can lead, for a nucleus with neutron
excess and ground state isospin $T$, to 3 different isospins ($T,
T\pm1$) in the daughter nucleus. As a consequence, GT$_-$ strength
distributions have significantly more structure and extend over a
larger energy interval than the GT$_+$ distributions, making their
theoretical reproduction more challenging. For example, the shell
model description of the GT$_-$ distributions in $^{54,56}$Fe with the
KB3 interaction failed \cite{Martinez95}, as the gross structure of
the distributions, experimentally observed at excitation energies
around 10 MeV, were noticeably shifted up in energy. These authors
pointed out that this discrepancy is related to the quasiparticle gap
around $N=28$ being too large. If this is indeed the case, then the
disagreement should be reduced by the slight monopole corrections in
the KB3 interaction defined above.  This conjecture is confirmed in
Fig.~\ref{fig2}, which shows the calculated GT$_-$ strength
distributions for $^{54,56}$Fe and $^{58,60}$Ni.

For $^{54}$Fe, the (p,n) cross sections have been converted into
GT$_-$ strength functions \cite{Anderson}, allowing a direct
comparison to our calculated distributions. Although not perfect in
details, the overall structure of the GT data are nicely reproduced by
our calculation. For the total GT$_-$ strength we find 6.9 (note that
the Ikeda sumrule includes also the universal scaling factor
$(0.74)^2$), which agrees well with the experimental results
($7.5\pm1.2$, $8\pm1.9$, $7..5\pm0.7$ \cite{Anderson,Rapaport}).

Ref.~\cite{Rapaport} reports the forward-angle (p,n) cross sections
for $^{54}$Fe, without conversion into GT strength functions. These
data are shown in the upper parts of Fig.~\ref{fig2}; they exhibit the
same peak structure as the GT data, but the conversion factor is
slightly energy dependent. The two data sets \cite{Anderson,Rapaport}
appear to have a small energy offset ($\approx 0.5 $ MeV), which might
be interpreted as the experimental energy uncertainty.

For $^{56}$Fe and $^{58,60}$Ni the forward-angle cross sections have
not been converted into GT$_-$ strength distributions. Therefore we
have not overlaid the data with the calculated strength distributions
in Figs.  2b-d, but rather show them in separate panels.  Our
calculations do not include the Fermi transitions to the isobaric
analog states which are therefore marked in the data.

For $^{56}$Fe the data show two major areas of GT strength, at
excitation energies below 4 MeV and between 6 and 15 MeV. This is well
reproduced with our modified interaction, just removing the overall
shift by 2 MeV encountered with the KB3 interaction. Although the
overall agreement is satisfying, the calculation misses in some
details.  For example, the structure in the data around 6 MeV is not
reproduced.  The calculated total GT$_-$ strength of 9.3 units agrees
with the experimental value of $9.9\pm2.4$ \cite{Rapaport}.

For $^{58,60}$Ni the major structures in the data are reproduced. But
we notice that the data for $^{58}$Ni exhibit structure at excitation
energies around 4 MeV, which is missing in our calculation. The
position and width of the major peak at 9 MeV, however, is nicely
reproduced.  For $^{60}$Ni the calculated GT$_-$ strength function
shows two narrow structures below 4 MeV, while the major strength
resides in a wide structure between 7 and 13 MeV. This is in good
agreement with the data. We calculate total GT$_-$ strengths of 7.7
units and 10.0 units for $^{58}$Ni and $^{60}$Ni, respectively, while
the experimental values are $7.4\pm1.8$ and $7.2\pm1.8$
\cite{Rapaport}. We note that applying the (scaled) Ikeda sumrule to
the (n,p) data for $^{60}$Ni, results in a GT$_-$ strength of 9.7
units,  larger than the value quoted in Ref. \cite{Rapaport}.

Although the centroids of the GT$_+$ strength distributions dominate
the stellar electron capture rates (and via the back-resonance also
the beta decay rates), weaker low-lying transitions can contribute due
to phase space enhancement. In the compilation, experimental
informations, derived from measured halflives, will be used, if
available. However, these are only available for ground states, while
excited states might contribute to the stellar rates via thermal
population. It is therefore useful to check how well our shell model
calculations do in comparison with measured halflives. Table II lists
halflives for selected nuclei, covering the mass range of interest.
These halflives have been calculated using the shell model matrix
elements and the experimental $Q_\beta$ values. We find that our
calculations agree usually very well with the data, a fact which has
already been noticed before in shell model calculations of lighter
nuclei (using the unmodified KB3 interaction, which yields very
similar results to the present modified version for pure $f_{7/2}$
nuclei).

The $Q_\beta$ values for beta decay increases significantly if one
moves to the proton- or neutronrich sides of a mass parabola. As
nuclei with large $Q_\beta$ values can, in general, decay to several
states of the daughter nucleus, the reproduction of such decay schemes
is a rather stringent test for nuclear models. Recently there has been
considerable interest in the properties of the double-magic nucleus
$^{56}$Ni and its neighbors. In this context, the $\beta_+$ decay of
the $^{56}$Cu ground state (with $J^\pi=4^+$) has been investigated
and the ft-values for the transitions to 3 final states with $J=4$ in
$^{56}$Ni could be determined \cite{Roeckl}.  Our shell model results
are compared to the data in Table III, and again good agreement is
observed.  To derive our results we have used the experimental
energies and adopted the $Q_{ec}$ value of 15.3 MeV \cite{Audi} from a
calculation of the Coulomb energy differences. We calculate a
$^{56}$Cu halflife of 67 ms, which agrees well with observation
($78\pm15$ ms) \cite{Roeckl}).

At stellar conditions electron capture and beta-decay also occur from
excited states. Although the compilation \cite{Langanke99} will use
the experimental excitation energies, it is nevertheless useful to
test how well our interaction reproduces the spectra of nuclei in the
mass range $A=45-65$. This will also give us an estimate of the
accuracy expected for very neutron-rich nuclei possibly encountered
during a core collapse and for which the spectral information at
modest energies is only incompletely known.

\begin{figure}
  \begin{center}
    \leavevmode
    \epsfxsize=0.8\columnwidth
    \epsffile{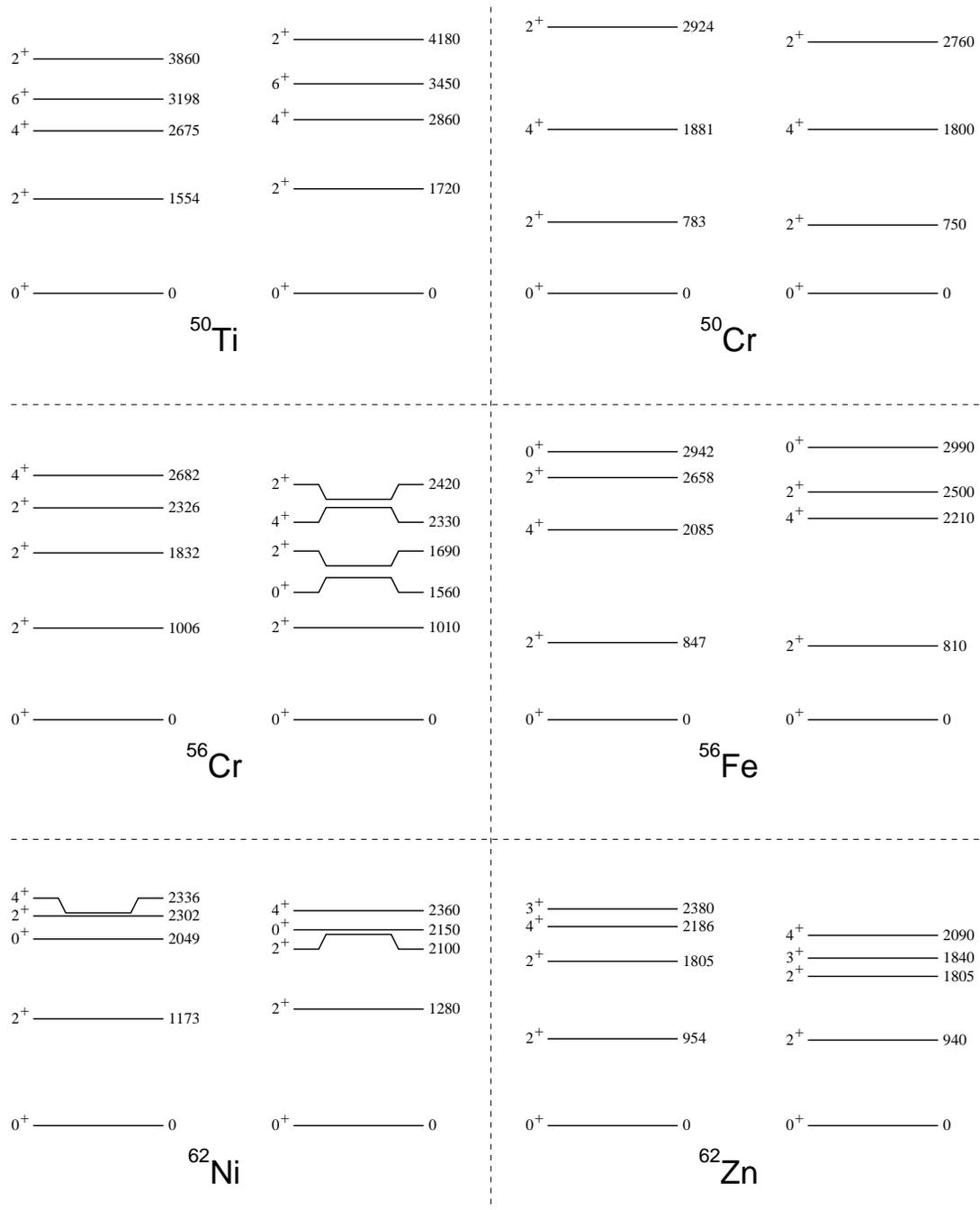}
    \caption{Comparison of experimental (left) and calculated (right)
      spectra for selected even-even nuclei. }
    \label{fig:even}
  \end{center}
\end{figure}

\begin{figure}
  \begin{center}
    \leavevmode
    \epsfxsize=0.8\columnwidth
    \epsffile{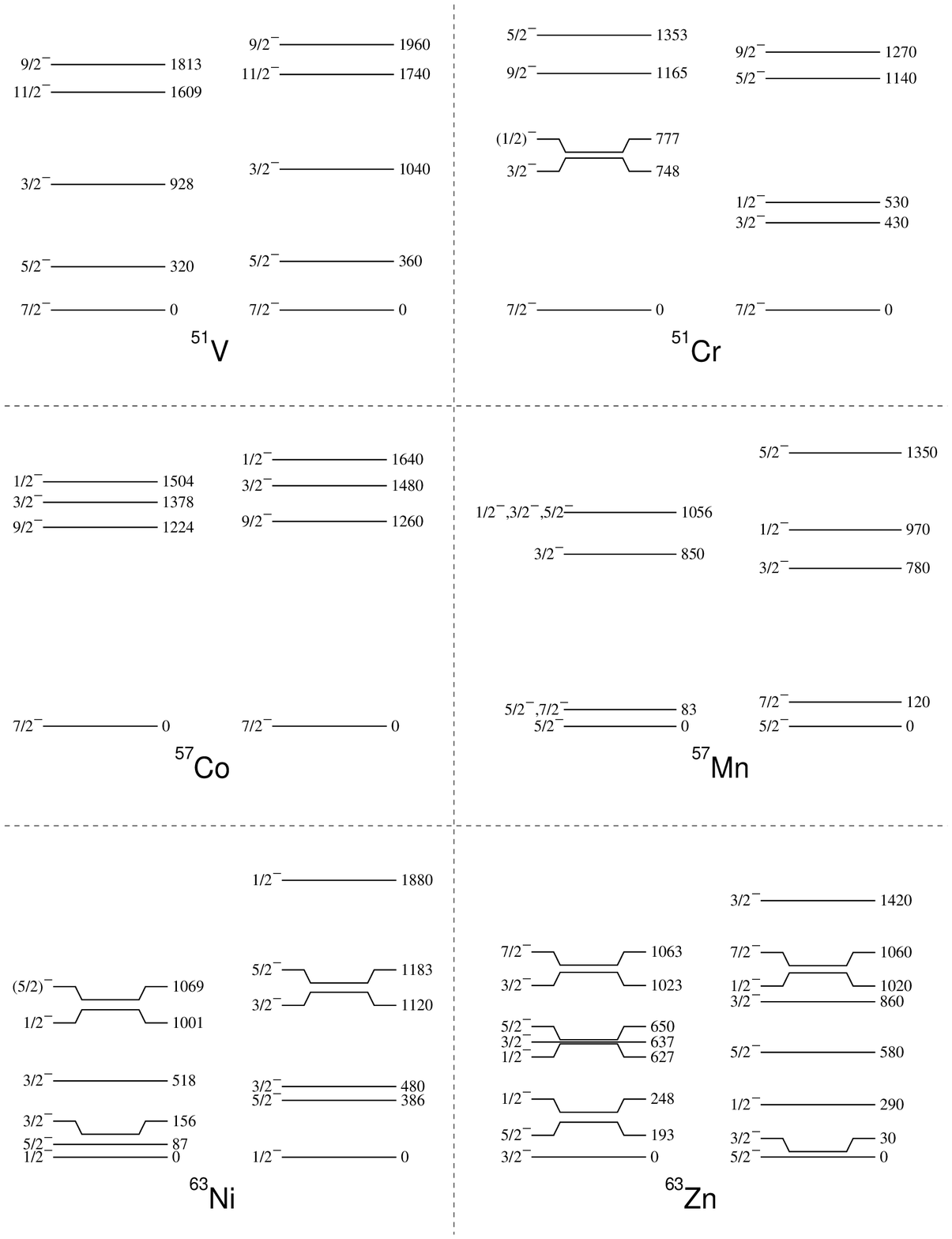}
    \caption{Comparison of experimental (left) and calculated (right)
      spectra for selected odd nuclei.}
    \label{fig:odd}
  \end{center}
\end{figure}

\begin{figure}
  \begin{center}
    \leavevmode
    \epsfxsize=0.7\columnwidth
    \epsffile{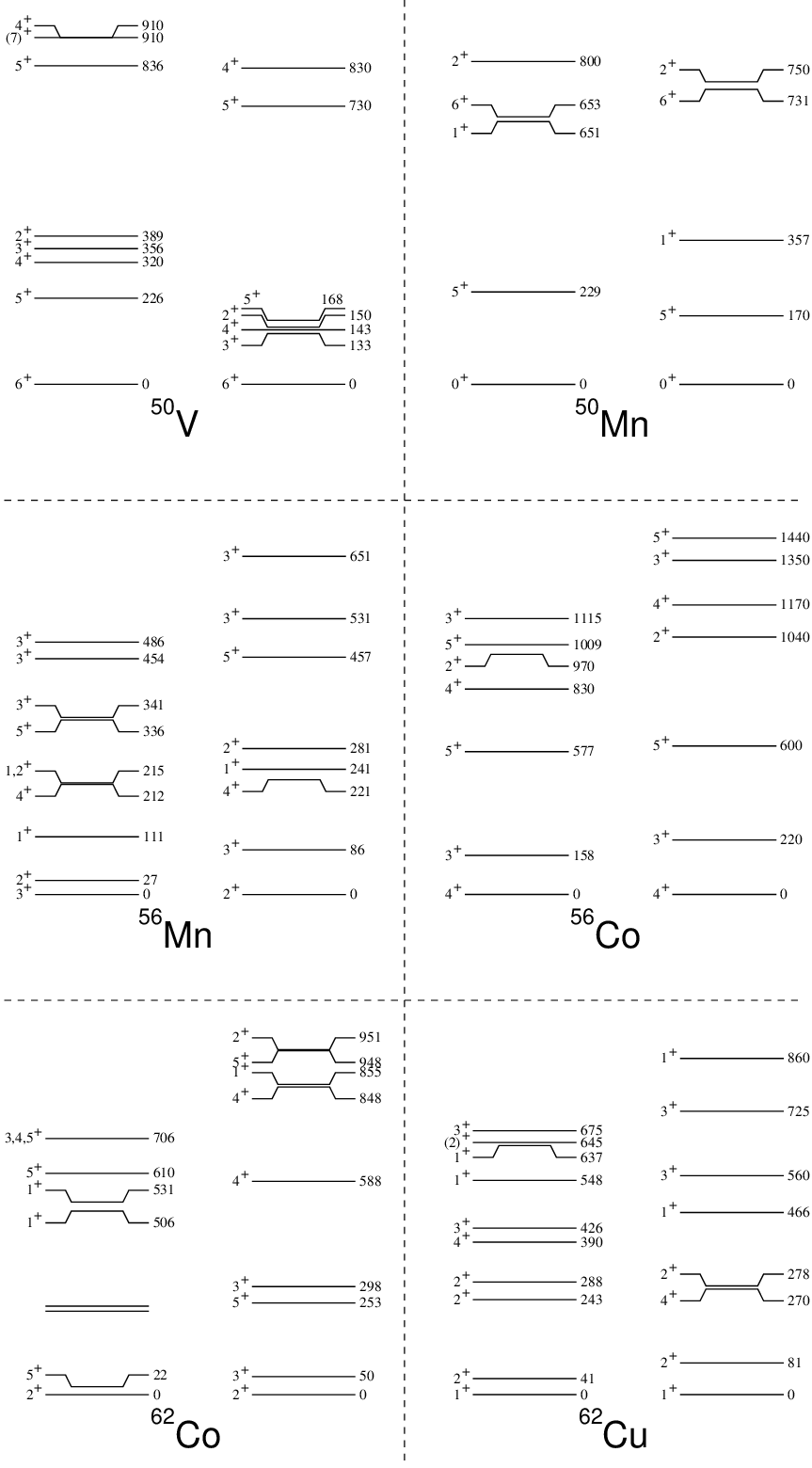}
    \caption{Comparison of experimental (left) and calculated (right)
      spectra for selected odd-odd nuclei.}
    \label{fig:odd-odd}
  \end{center}
\end{figure}

We compare the calculated spectra with data for selected
nuclei covering the relevant mass range. The presentation is for
even-even (fig.~\ref{fig:even}), odd-$A$ (fig.~\ref{fig:odd}) and
odd-odd (fig.~\ref{fig:odd-odd}) nuclei separately.  Furthermore we
compare several nuclei of the same mass number $A$ to test the isospin
dependence of the interaction. We note that at the chosen truncation
levels the energies are not completely converged yet, which might
result in slight inaccuracies if judging the validity of the
interaction.  Nevertheless we chose to present the results consistent
with the shell model evaluation of the stellar weak interaction rates
as we feel it is more important to know the accuracy of the
calculations behind these astrophysically important quantities.

For even-even nuclei the low-lying spectrum comprises mainly the $0^+$
ground state, the two lowest $2^+$ states and the first excited $4^+$
state.  We see that these states are well reproduced in the relevant
mass range. For odd-$A$ nuclei the spectra are generally also well
described, although we note that the interaction is not perfect for
nuclei with masses larger than $A=60$.  For example, the two lowest
states in $^{63}$Zn are inverted in our calculation (although the
misplacement amounts only to about 200 keV).  Obviously the most
difficult spectra to describe are those of odd-odd nuclei. Again the
interaction does pretty well, although it inverts, for example, the
two lowest states in $^{56}$Mn. Note, however, that on an absolute
scale these misplacements are rather minute and amount only to 130
keV.  For $^{62}$Cu, three $2^+$ and two $1^+$ and $3^+$ levels have
been experimentally idenfied below 700 keV excitation energy; our
calculation only misses one $2^+$ state. Experimentally the $^{62}$Co
spectrum is incompletely known. For example no spin assignment has
been made for the states at 230 keV and at 244 keV. Our calculation
predicts two $3^+$ states around these energies.  The spectra of the
nuclei in the mass range $A=45-65$, which are not shown here, are of
similar quality than those presented and discussed.  In particular the
reproduction of the observed level schemes by our interaction gets
slightly worse if moving to larger mass numbers within the $pf$ shell.

In summary, our  goal is to calculate stellar weak interaction
rates as they are needed for example in simulations of the core
collapse of a massive star. The method of choice for this adventure is
the interacting shell model and in this manuscript we aimed at
demonstrating that modern large-scale shell model calculations have
the predictive power to fulfill this job. To demonstrate this we have
performed large-scale shell model calculations for nuclei in the mass
range $A=45-65$ using a slightly modified version of the wellknown KB3
interaction in which we have corrected small inefficiencies of the
monopole terms.  After these modifications the KB3 interaction is
suited for applications in shell model calculations of stellar
electron capture and beta decay rates. Our studies reproduce those
ingredients which have been identified to be essential for a reliable
estimate of the stellar rates. These are most notably the $GT_+$
strength distributions in the daughter nucleus (for nuclei with
$N>Z$), which are well reproduced in all cases for which
experimentally data are available.  This includes also rather weak GT
transitions as they usually determine the halflives of nuclei in this
mass range under laboratory conditions. Beta decay rates also depend
on the $GT_-$ strength distributions and again our calculations
reproduce the available experimental data on $^{54,56}$Fe and
$^{58,60}$Ni rather well, clearly improving previous studies. Finally
we have demonstrated that the shell model calculations give a fair
account of the level spectra of nuclei in this mass regime. This is of
interest for the calculations of stellar weak interaction rates as
they occur at finite temperature.

Summarizing, for the first time a theoretical tool is in hand which
has the predictive power to reliably calculate the stellar weak
interaction rates for nuclei in the mass range $A=45-65$.  These
computationally intensive studies are in progress and their results
will be available soon.

\acknowledgements

The authors are very grateful to Professor Alfredo Poves for his
advices and many useful discussions.  This work was supported in part
by the Danish Research Council.

\end{multicols}

\newpage

\begin{table}
\begin{center}
  \caption{Comparison of the calculated and experimental total GT$_+$
    strengths.  The data are from
    \protect\cite{gtdata1,gtdata2,gtdata3,gtdata4,gtdata5} The
    calculated centroids of the GT$_+$ distributions are compared to
    the GT resonance energy assumed in the FFN parametrization; both
    energies are in MeV.}
  \begin{tabular}{ccccc}
    nucleus & experiment &shell model  &centroid &FFN\\
    \hline
    $^{51}$V &    1.2$\pm$0.1 & 1.4 & 5.18 & 3.83\\
    $^{54}$Fe&    3.3$\pm$0.5 & 3.6 & 3.78 & 3.80\\
    $^{55}$Mn&    1.7$\pm$0.2 & 2.2 & 4.57 & 3.79\\
    $^{56}$Fe&    2.8$\pm$0.3 & 2.7 & 2.60 & 3.78\\
    $^{58}$Ni&    3.8$\pm$0.4 & 4.4 & 3.75 & 3.76\\
    $^{59}$Co&    1.9$\pm$0.1 & 2.5 & 5.05 & 2.00\\
    $^{60}$Ni&    3.1$\pm$0.1 & 3.4 & 2.88 & 2.00\\
    $^{62}$Ni&    2.5$\pm$0.1 & 2.1 & 1.78 & 2.00\\
    $^{64}$Ni&    1.7$\pm$0.2 & 1.3 & 0.50 & 2.00\\
  \end{tabular}
\end{center}
\end{table}

\begin{table}
  \begin{center}
    \caption{Comparison of $\beta_+$ halflives with the shell model
      results for selected nuclei.  }
    \begin{tabular}{lcc}
      Nuclei & Expt.  &Shell Model  \\
      \hline
      $^{51}$Cr  & 27.7 d & 27 d  \\
      $^{53}$Fe  & 8.51 m &7.0 m \\
      $^{54}$Mn  & 312 d  &243 d  \\
      $^{55}$Co  & 17.5 h &16.7 h  \\
      $^{57}$Co  & 272 d  &397 d  \\
      $^{57}$Ni  & 35.6 h &49 h   \\
      $^{58}$Co  & 70.8 d & 50.8 d   \\
      $^{60}$Cu  & 23.7 m &13.5 m   \\
      $^{61}$Cu  & 3.3 h  &5.0 h   \\
    \end{tabular}
  \end{center}
\end{table}

\begin{table}
  \begin{center}
    \caption{Comparison of the partial $\beta_+$ decay of $^{56}$Cu
      with the shell model results. The energies are in MeV.  }
    \begin{tabular}{lcccc}
      final state & energy (exp) & energy (SM)& log ft (exp) & log ft 
      (SM)  \\
      \hline
      $4^+$, T=0 & 3.926 & 4.495 & 4.2(2) & 4.38 \\
      ($3^+$, T=0)&       & 4.902 &        & 4.80 \\  
      $4^+$, T=0 & 5.483 & 5.893 & 4.2(2) & 4.33 \\
      $4^+$, T=1 & 6.432 & 6.113 & 3.6(1) & 3.49 \\
      ($3^+$, T=1)&       & 6.335 &        & 4.28 \\          
    \end{tabular}
  \end{center}
\end{table}

\end{document}